\documentclass[twocolumn]{aastex7}

\usepackage{amsmath}	% Advanced maths commands
\usepackage{amssymb}	% Extra maths symbols
\usepackage{xcolor}
\usepackage{enumitem}
\usepackage{tikz}
\usepackage{hyperref}
\newcommand{\JA}{\hyperlink{ac:JA20}{JA20} }
\newcommand{\JAnospace}{\hyperlink{ac:JA20}{JA20}}
\newcommand{\JApar}{(\hyperlink{ac:JA20}{JA20})}

\newcommand{\dd}{{\rm d}}
\newcommand{\be}{\begin{equation}}
\newcommand{\ee}{\end{equation}}
\newcommand{\units}[1]{\ensuremath{\, \mathrm{#1}}}  % Shorthand for proper spacing with units.

\begin{document}

\title{Stellar separation shapes spin-orbit alignment in visual binaries}
\shorttitle{Spin-orbit alignment in visual binaries}
\shortauthors{Poon \& Pham et al.}

\author[0000-0001-7739-9767]{Michael Poon}
\altaffiliation{MP and DP contributed equally and are joint first authors.}
\affiliation{Department of Astronomy and Astrophysics, University of Toronto, Toronto, Ontario, M5S 3H4, Canada}
\email[show]{michael.poon@astro.utoronto.ca}

\author[0000-0002-0924-8403]{Dang Pham}
\altaffiliation{MP and DP contributed equally and are joint first authors.}
\affiliation{JILA and Department of Astrophysical and Planetary Sciences, University of Colorado Boulder, Boulder, CO 80309, USA}
\email[show]{dang.pham@colorado.edu}

\author[0000-0002-6076-5967]{Marta L. Bryan}
\affiliation{Department of Astronomy and Astrophysics, Penn State University, 525 Davey Laboratory, 251 Pollock Road, University Park, PA 16802, USA}
\affiliation{Department of Chemical and Physical Sciences, University of Toronto Mississauga, Mississauga, Ontario, L5L 1C6, Canada}
\affiliation{Department of Astronomy and Astrophysics, University of Toronto, Toronto, Ontario, M5S 3H4, Canada}
\email{mlb645@psu.edu}

\author[0000-0003-1927-731X]{Hanno Rein}
\affiliation{Department of Physical and Environmental Sciences, University of Toronto Scarborough, Toronto, Ontario, M1C 1A4, Canada}
\affiliation{Department of Astronomy and Astrophysics, University of Toronto, Toronto, Ontario, M5S 3H4, Canada}
\email{hanno.rein@utoronto.ca}

\author[0000-0002-3610-6953]{Jiayin Dong}
\affiliation{Department of Astronomy, University of Illinois at Urbana-Champaign, Urbana, IL 61801, USA}
\affiliation{Center for Astrophysical Surveys, National Center for Supercomputing Applications, Urbana, IL 61801, USA}
\email{jdongx@illinois.edu}

\begin{abstract}

Stellar binaries may form through several formation pathways, including disk or core fragmentation. Their spin-orbit angles are a signature of formation, although individual measurements for visual binaries are limited and broad. A seminal work by \citet{Hale1994} found that visual binaries with separations $\lesssim 30$ AU tend to be more aligned, which laid the groundwork for binary formation theories. However, \citet{Justesen+Albrecht2020} found that underestimated stellar radii lead to inaccurate spin-orbit angles and that KS statistics do not provide meaningful population-level constraints even with updated radii. Using a hierarchical Bayesian model to reanalyze their dataset, we find evidence with a Bayes factor of 12 for two subpopulations of spin-orbit angles separated by a $\sim 31-38$ AU cutoff. Binaries inside (outside) the cutoff are more (less) aligned, consistent with a Fisher distribution with $\kappa=48$ ($\kappa=6$). We also find possible indications of a secondary cutoff at $\sim 10-17$ AU, although more data is required to resolve this prediction. 
These cutoffs may mark transitions between formation pathways: closer-in binaries tend to form aligned in a shared protostellar disk, while wider binaries tend to form less aligned through turbulent fragmentation.

\end{abstract}

\keywords{Visual binary stars (1777), Hierarchical models (1925), Astrostatistics (1882)}

\section{Introduction} \label{sec:intro}

In stellar binaries, the angle between a star's spin axis and orbital axis -- the spin-orbit angle -- is a tracer of their formation history \citep{Bate2012, Smith+2024}. Stellar binaries that obtain their spin and orbital angular momentum through formation in a shared protostellar disk are expected to exhibit spin-orbit alignment, while turbulent formation triggered by gravitational instability can imprint a wide range of spin-orbit angles \citep{Offner+2016}. 

An early landmark study by \citet{Hale1994} found that binary stars tend to have aligned spin-orbit angles for semi-major axes $\lesssim 30-40\units{AU}$, and randomly-oriented angles at wider separations. This result has become foundational to our understanding of binary formation and evolution \citep{Mathieu1994, Bate2009, Bate2018}. 

Decades later, a follow-up study by \citet{Justesen+Albrecht2020}
(hereafter \hypertarget{ac:JA20}{JA20}) casts doubt on the existence of this semi-major axis cutoff, primarily due to an underestimation of stellar radii in \citet{Hale1994}, leading to less reliable spin-orbit angle measurements. With updated stellar radii from high-resolution spectra and Gaia DR2, \JA finds a different result than \citet{Hale1994}. Using Kolmogorov-Smirnov (KS) statistics, \JA cannot reject either hypotheses that their data came from random or perfectly aligned populations.
That is, they find inconclusive evidence for an isotropic or perfectly-aligned distribution of stellar spin-orbit angles, potentially due to broad uncertainties.
If there is indeed no evidence for the $30-40\units{AU}$ cutoff, this would call into question a widely cited observational constraint on binary formation models.

In this study, we reanalyze the \JA dataset using hierarchical Bayesian modeling to determine whether evidence exists for a $30-40\units{AU}$-cutoff.
Similar to \JAnospace, we find the spin-orbit angle posterior distribution for each star.
Then, instead of the KS test, we implement a hierarchical Bayesian model to infer the population-level distribution of spin-orbit angles and determine if there is a cutoff.
In this approach, the population(s) of spin-orbit angles do not have to be either isotropic or perfectly aligned, but can rather be described through a parameterized distribution.

We develop two statistical models in Sec. \ref{sec:model}, based on the hierarchical Bayesian model in \citet{Poon+2025} for planetary spin-orbit angles.
The first model describes all stellar spin-orbit angles with a single population distribution, while the second model allows the distribution to differ on either side of a cutoff separation. In Sec. \ref{sec:results}, we apply these models to the \JA dataset and answer the following statistical questions:
\begin{itemize}
    \item If stellar spin-orbit angles come from the same population, how is that single population described? (Sec. \ref{sec:results_single})
    \item Is a two-population model preferred over a single population? (Sec. \ref{sec:results_Bayes_factor})
    \item If there are two populations, how are they described? (Sec. \ref{sec:results_kappa})
    \item Could there be more than two populations? (Sec. \ref{sec:results_double_cutoffs})
\end{itemize}
In Sec. \ref{sec:discussion}, we compare and contrast our results to previous works and discuss these findings in the broader stellar binary context.
In Sec. \ref{sec:conclusions}, we summarize our findings.

\section{Statistical Modeling} \label{sec:model} 

\begin{figure}
    \centering
    \includegraphics[width=0.8\linewidth]{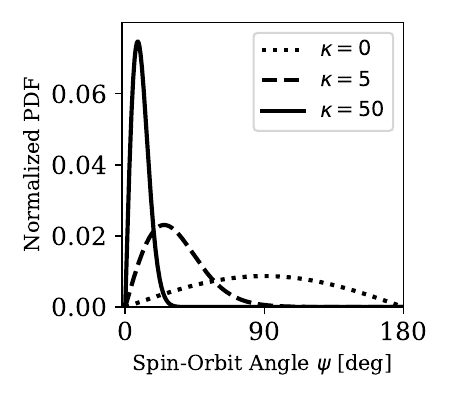}
    \caption{Fisher distribution for several values of $\kappa$, showing the resulting spin-orbit angle $\psi$ probability density functions. Small $\kappa$ values produce nearly isotropic distributions, whereas large $\kappa$ values indicate strong alignment.
    }
    \label{fig:kappa}
\end{figure}

\begin{figure}
    \centering
    \includegraphics[width=1.\linewidth]{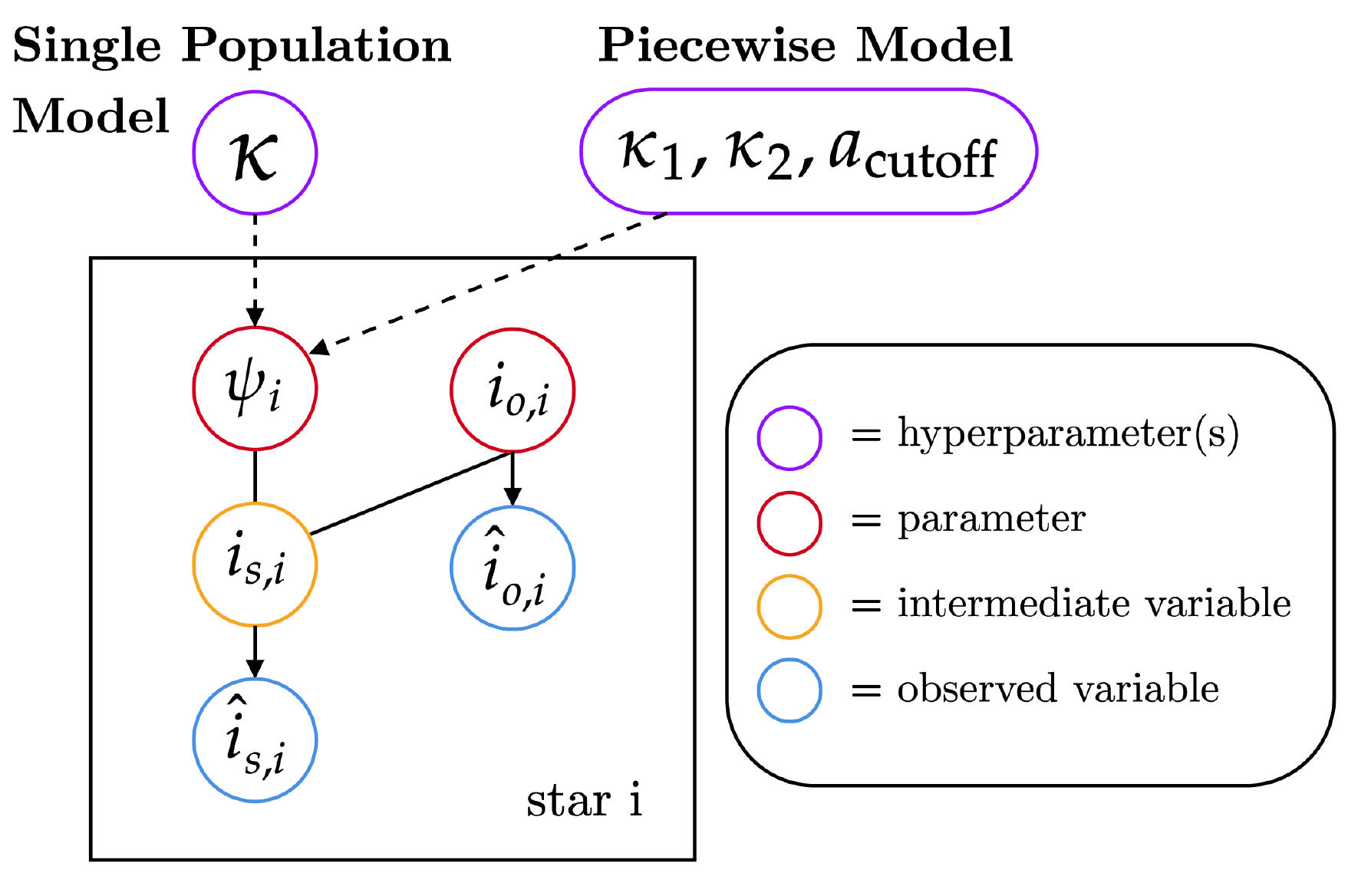}
    \caption{Graphical representation of our two hierarchical models. Arrows represent the direction of data generation either from the Single Population Model or the Piecewise Model to $\psi_i$, the spin-orbit angle of star $i$, to the observed variables $\hat{i}_s$ (stellar spin axis inclination) and $\hat{i}_o$ (orbital inclination).
    }
    \label{fig:plate_diagram}
\end{figure}

Individual measurements of spin-orbit angles in visual binaries are difficult to interpret because only the line-of-sight component of the true angle in 3D space is observable \JApar. As a result, we can only determine a lower limit on their intrinsic alignment or misalignment. With sufficient data, population-level trends in these projected angles can distinguish between truly aligned and misaligned ensembles.

The seminal work of \citet{Hale1994} analyzed 45 solar-type binaries, and found that binaries with separations $< 20\units{AU}$ exhibit small ($<20^\circ$) line-of-sight projected spin-orbit angles, while wide binaries with separation $> 30\units{AU}$ exhibit a wide range of misalignment up to $\sim 60^\circ$, consistent with random spin-orbit orientations. In recent follow-up analysis, \JA more accurately determine stellar radii from Gaia parallaxes, and reveal that some binaries $< 20\units{AU}$ in the \citet{Hale1994} instead have greater misalignment.
The \JA dataset (cf. their Table 1) has 33 stars with semi-major axes ranging from $\sim 3-300\units{AU}$.
In this work we reanalyze the same \JA dataset using a different statistical technique: hierarchical Bayesian modeling.

Our hierarchical Bayesian model is based on the recent work by \citet{Poon+2025}, who used this methodology to infer the spin-orbit angle distribution of super-Jupiters.
Their inference relies on nearly the same observables as in visual binaries, except that planetary rotation is measured instead of stellar rotation. We therefore build on the \citet{Poon+2025} framework to test the foundational result of \citet{Hale1994}: that visual binary separation shapes their spin-orbit alignment.
While \citet{Poon+2025} assumes that all spin-orbit angles originate from a single population, we extend their model to allow spin-orbit angles to be drawn from different populations depending on their binary separation.
In this section, we summarize the framework and highlight the key modifications required to analyze populations of stellar spin-orbit angles.

\subsection{Single Population Model}\label{sec:model_single}

For a given star, its spin axis inclination $i_s$ and orbital inclination $i_o$ can be constrained through a combination of astrometry and high-resolution spectroscopy \citep{Masuda+Winn2020, Justesen+Albrecht2020}. Combining $i_s$ and $i_o$ in a Bayesian framework yields the posterior of the spin-orbit angle $\psi$ (see Eqn. 2 of \citealt{Poon+2025}). This posterior $\mathcal{P}(\psi | i_s, i_o)$ requires a prior $\pi(\psi)$ and likelihood $\mathcal{L}(i_s | \psi, i_o)$.

The prior for $\psi$ for each star is described by a Fisher distribution \citep{Fisher1953}:
\begin{equation}
    \pi(\psi|\kappa) = \frac{\kappa}{2\sinh{\kappa}}\exp{(\kappa \cos{\psi})}\sin{\psi},
\label{eq:fisher}
\end{equation}
which depends on a concentration parameter $\kappa$. As shown in Fig. \ref{fig:kappa}, increasing $\kappa$ concentrates the spin-orbit angle distribution towards alignment, while decreasing $\kappa$ produces a more random orientation.
The prior is conditioned on a given $\kappa$, which we draw from a hyperprior in the hierarchical Bayesian model. For the single population model, we choose a hyperprior for $\kappa$ following \citet{Munoz+Perets2018}:
\begin{equation}
    \mathcal{H}(\kappa) \propto (1 + \kappa^2)^{-1/2},
\label{eq:prior}
\end{equation}
which avoids strong preference towards perfect alignment or random orientation. The likelihood for $\mathcal{L}(i_s | \psi, i_o)$ is derived in Eqn. 11 of \citealt{Poon+2024a}.

Following \citet{Poon+2025}, our hierarchical structure is summarized in Fig. \ref{fig:plate_diagram}, where we denote the true and observed quantities by $(x, \hat{x})$, and use $\{x_i\}=\{x_i\}_{i=1}^n$ for a population of $n$ independent stars. Combining the hyperprior, prior, and likelihood, while marginalizing over individual systems quantities $\psi_i$ and $i_{o,i}$, produces the following posterior:

\begin{widetext}
\begin{equation}
    \begin{split}
    \mathcal{P}\left(\kappa \Big|\{\hat{i}_{s,i},\hat{i}_{o,i}\}\right) \propto \underbrace{\mathcal{H}(\kappa)}_{\text{hyperprior}} \times
  \underbrace{\prod_{i=1}^n}_{\text{iid}} \int \underbrace{\pi(\{\psi_i, i_{o,i}\}|\kappa)}_\mathrm{prior} ~\times~\underbrace{\mathcal{L}(\{\hat{i}_{s,i},\hat{i}_{o,i}\}|\{\psi_i, i_{o,i}\})}_\mathrm{likelihood} ~\dd \psi_i ~\dd i_{o,i}.
\label{eq:posterior2}
\end{split}
\end{equation}
\end{widetext}

For full details and model performance tests, refer to Section 2 and Appendix B in \citet{Poon+2025}. We have now constructed a hierarchical model that infers $\kappa$, assuming a single population. Next, we develop a piecewise model that infers $\kappa$ for sub-populations.

\subsection{Piecewise Model}\label{sec:model_piecewise}

In this model, stellar spin-orbit angles are drawn from one of two distributions (with different $\kappa$), depending on whether the binary separation is above or below a cutoff semi-major axis $a_\mathrm{cutoff}$.
We implement this by modifying the posterior in Eqn. \ref{eq:posterior2} as follows.
The choice of $\kappa$ in the prior $\pi(\{\psi_i, i_{o,i}\}|\kappa)$ now takes the form of a piecewise function:
\begin{equation}\label{eq:kappa}
    \kappa= \begin{cases} 
        \kappa_1, & a\leq a_\mathrm{cutoff} \\
        \kappa_2, & a > a_\mathrm{cutoff}
    \end{cases}
\end{equation}
where $\kappa_1$ and $\kappa_2$ describe the two populations' spin-orbit angle distribution.
In this formulism, $\kappa_1$, $\kappa_2$, and $a_\mathrm{cutoff}$ are independent random variables.

While $\kappa_1$ and $\kappa_2$ are continuous random variables, ranging from $[0, \infty)$, $a_\mathrm{cutoff}$ is a discrete random variable. For a sample of $N$ stellar binaries semimajor axes, there are only $N-1$ possible values for $a_\mathrm{cutoff}$. For example, when $N=2$, the only possible cutoff is between the two distinct semimajor axes, and all values between these bounds have equal probability.

The hyperprior now depends on these random variables, such that 
\begin{equation}\label{eqn:hyperprior}
    \mathcal{H}=P(\kappa_1) P(\kappa_2)P(a_\mathrm{cutoff}).
\end{equation}
Here, $P(\kappa_1)$ and $P(\kappa_2)$ take on the same functional form of Eqn. \ref{eq:prior}. 
The choices for $P(a_\mathrm{cutoff})$ will be discussed in Sec. \ref{sec:results_Bayes_factor}.

The likelihood probability function remains the same, as it is not conditioned on $\kappa$.
Thus, the posterior distribution for the piecewise model\footnote{The piecewise model can be viewed as a special case of a more general mixture model. These two formulations yield consistent posteriors for $\kappa_1$ and $\kappa_2$, as shown in Appendix \ref{appendix:mixture}.} takes the form $\mathcal{P}\left(\kappa_1, \kappa_2, a_\mathrm{cutoff} \Big|\{\hat{i}_{s,i},\hat{i}_{o,i}\}\right)$.

\section{Results}\label{sec:results}

With the single population and piecewise models constructed, we now answer the previously posed statistical questions in Sec. \ref{sec:intro}\footnote{The \JA dataset does not include uncertainties in the orbital inclination, $i_o$. In Appendix \ref{appendix:sensitivity}, we show that adding $1^\circ-5^\circ$ uncertainties to $i_o$ does not change, but does reduce the strength of our conclusions.}.

\subsection{Single Population}\label{sec:results_single}

\begin{figure}
    \centering
    \includegraphics[width=1\linewidth]{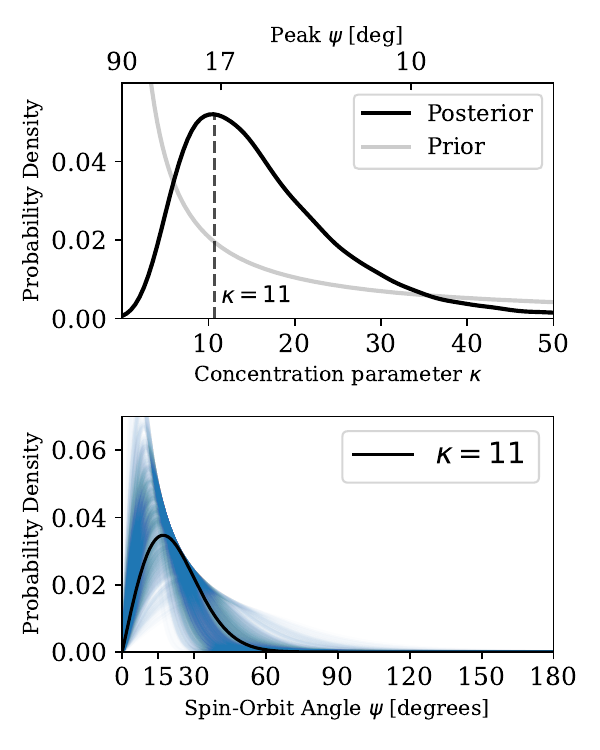}
    \caption{\textbf{Top}: Posterior distribution of $\kappa$ for the Single Population Model. The mode and 68\% highest density probability interval is $\kappa = 11\substack{+10 \\ -6}$, corresponding to spin-orbit angle distributions peaking at $\psi = 17\substack{+8 \\ -5}\units{^\circ}$. \textbf{Bottom}: 1000 spin-orbit angle distributions shown as blue curves, each corresponding to a sample from the $\kappa$ posterior. In black is the $\psi$ distribution for $\kappa=11$ which peaks at $17^\circ$, signifying moderate spin-orbit alignment.
    }
    \label{fig:single_kappa}
\end{figure}

We sample the posterior distribution for $\kappa$ (Eqn. \ref{eq:posterior2}) using \texttt{emcee} \citep{ForemanMackey2013}, which has been demonstrated to be excellent at sampling posteriors in a wide variety of statistical contexts.
In Fig. \ref{fig:single_kappa}, we present the results for $\kappa$, if all stellar spin-orbit angles are drawn from one distribution (Sec. \ref{sec:model_single}).
We find that $\kappa$ is a positively-skewed distribution, with a peak and the 68\% highest density probability interval of $\kappa = 11\substack{+10 \\ -6}$.
This indicates that the distribution of spin-orbit angles for a single population have a peak at $\psi \sim 17 ^{\circ}$ (bottom panel of Fig. \ref{fig:single_kappa}).

\subsection{Is there a preference between a single population vs. piecewise model?}\label{sec:results_Bayes_factor}

\begin{figure}
    \centering
    \includegraphics[width=1.0\linewidth]{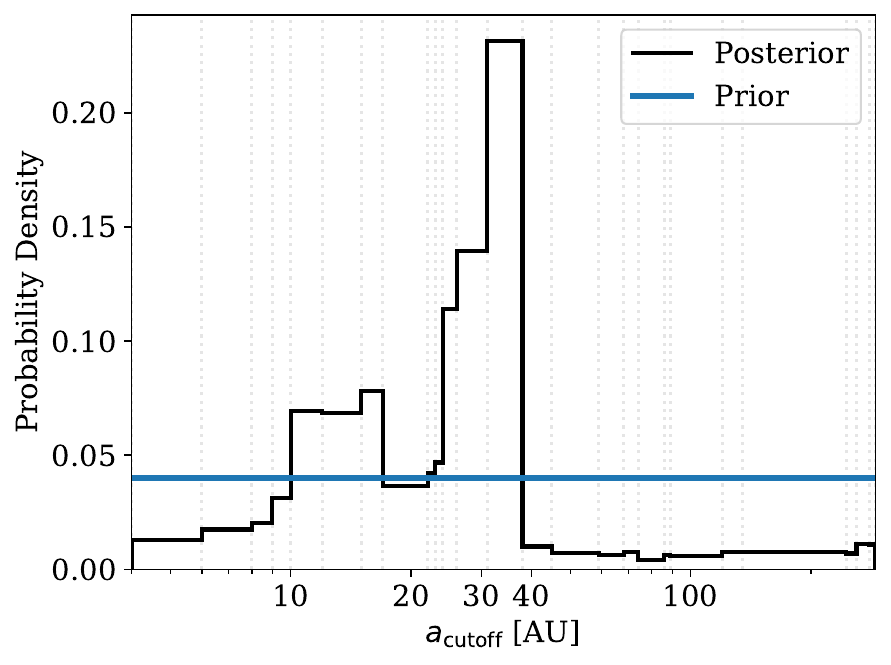}
    \caption{Posterior distribution of the semi-major axis cutoff ($a_\mathrm{cutoff}$) for the piecewise model, where vertical dotted lines represent the semi-major axis of each stellar binary. The prior is a uniform distribution. The region with the highest probability is $a_\mathrm{cutoff}\sim 31-38\units{AU}$.
    }
    \label{fig:a_posterior}
\end{figure}

Next, we investigate the piecewise model scenario, where the population is split into two, demarcated by a semi-major axis cutoff.
We infer the semi-major axis cutoff $a_\mathrm{cutoff}$ by adopting an uninformative prior such that $P(a_\mathrm{cutoff})$ (cf. Eqn. \ref{eqn:hyperprior}) is a uniform distribution.
In Fig. \ref{fig:a_posterior}, we show the posterior $\mathcal{P}\left(a_\mathrm{cutoff} \Big|\{\hat{i}_{s,i},\hat{i}_{o,i}\}\right)$.
We find that $a_\mathrm{cutoff} \sim 31-38$ AU have the highest probability, similar to the conclusion by \citet{Hale1994}.

However, having the highest probability does not mean that the piecewise model at $a_\mathrm{cutoff} \sim 31-38$ AU is preferred over the single population model.
To address this issue, we use the Bayes factor to test the piecewise model (at a given $a_\mathrm{cutoff}$) against the single population model.
First, we must modify the prior for $a_\mathrm{cutoff}$, such that $P(a_\mathrm{cutoff}) = \delta(a_\mathrm{cutoff})$, where $\delta(a_\mathrm{cutoff})$ is the Dirac delta function evaluated at the midpoint between stellar binary semi-major axes (see Sec. \ref{sec:model_piecewise} for a discussion regarding $a_\mathrm{cutoff}$ being a discrete random variable).
Now, we can sequentially test the relative preference of the piecewise model versus the single population model for various $a_\mathrm{cutoff}$, using the Bayes factor, defined as:
\begin{equation}
    BF_\mathrm{piecewise, single~pop.}(a_\mathrm{cutoff}) = \frac{Z_\mathrm{piecewise}(a_\mathrm{cutoff})}{Z_\mathrm{single}}
\end{equation}
where $Z$ is the evidence for a particular model, with full forms of $Z$ presented in Appendix \ref{appendix:evidence}.
This methodology allows the comparison between the piecewise model at various cutoff thresholds against the single population model.
A similar procedure is done in \citet{Poon+2025}, where $\kappa$ is held fixed, to test preferences between various $\kappa$ models.
We use nested sampling with \texttt{dynesty} \citep{Speagle2020} to calculate evidence for the two models. 
Nested sampling is more appropriate for evidence sampling, as compared to \texttt{emcee} which is best for posterior sampling.
\begin{figure}
    \centering
    \includegraphics[width=1.0\linewidth]{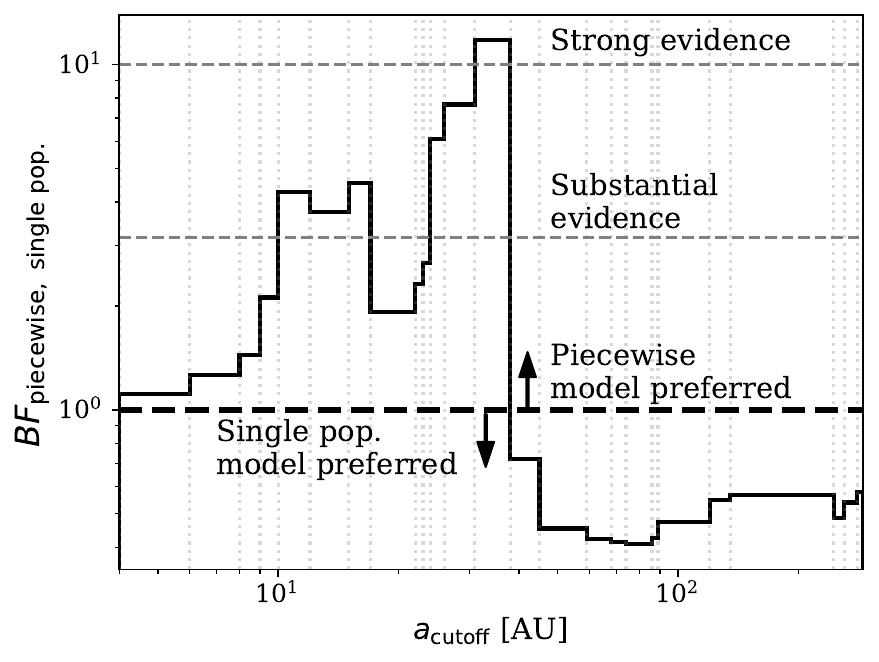}
    \caption{The Bayes factor ($BF_\mathrm{piecewise,~single~pop.}$) comparing the piecewise and single population models. The evidence for the piecewise is evaluated as a function of semi-major axis cutoff ($a_\mathrm{cutoff}$), where vertical dotted lines represent the semi-major axis of each stellar binary. Horizontal dashed lines represent the Bayes factor interpretation by \citet{Jeffreys1939}. A Bayes factor greater than 1 favors the piecewise model and vice-versa. The region with the highest evidence supporting the piecewise model is $\sim 31-38\units{AU}$ (strong evidence with Bayes factor of 12).
    }
    \label{fig:bayes_factor}
\end{figure}

In Fig. \ref{fig:bayes_factor}, we show the Bayes factor comparing the single population and piecewise models.
In the same figure, we interpret the Bayes factor using the scale proposed by \citet{Jeffreys1939}\footnote{The interpretations of our results remain the same when using an alternative Bayes factor scale by \citet{KassRaftery1995}.}.
We find that the \JA data (with a modest sample size of 33 stars) supports the piecewise model with a cutoff at $a_\mathrm{cutoff} \approx 31-38$ AU with strong evidence as the Bayes factor reaches $\approx 12$ in this region.
This is consistent with the original hypothesis by \citet{Hale1994}, where the author suggests a cutoff near $30$ AU.
Notably, we find that after the cutoff at $\sim 38$ AU, no conclusive model can be chosen.
In addition, there is another Bayes factor peak at $\sim 10-17$ AU, though with lower evidence.
As this could potentially imply a model with two cutoffs (i.e., stellar spin-orbit angles are drawn from three populations), we revisit this possibility in Sec. \ref{sec:results_double_cutoffs}.

\subsection{What are the two populations?}\label{sec:results_kappa}

\begin{figure*}
    \centering
    \includegraphics[width=1\linewidth]{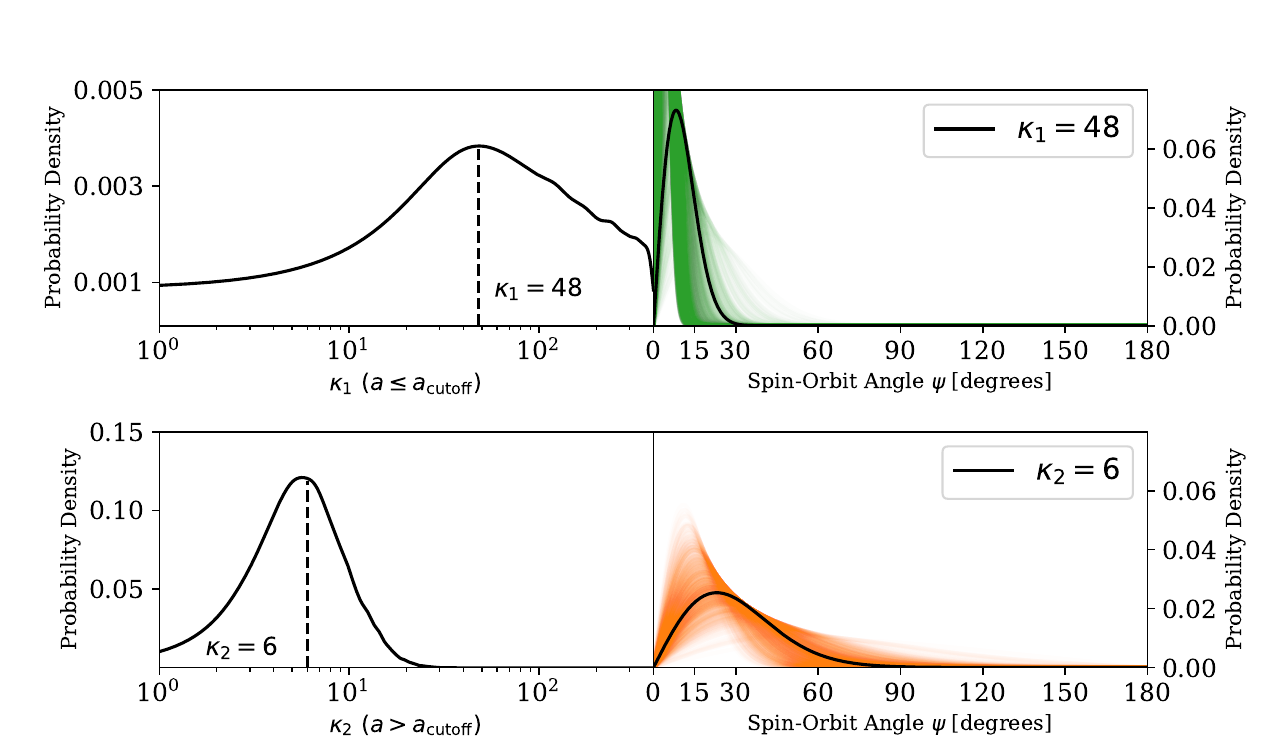}
    \caption{Posteriors for the two population model.
    \textbf{Top left}: The posterior for $\kappa_1$ (subpopulation with semi-major axis $a \leq a_\mathrm{cutoff}$). The mode and 68\% highest density probability interval (HDPI) is $\kappa_1 = 48_{-32}^{+194}$, corresponding to spin-orbit angles peaking at $\psi = 8\substack{+6 \\ -4}\units{^\circ}$. \textbf{Top right}: 1000 $\psi$ distributions shown as green curves, each corresponding to samples from the $\kappa_1$ posterior. The black curve is the $\psi$ distribution for $\kappa_1 = 48$, peaking at $10^{\circ}$ indicating highly-aligned spin-orbit angles.
    \textbf{Bottom}: Same as the top two panels, but for $\kappa_2$ (when $a > a_\mathrm{cutoff}$). The mode and 68\% HDPI is $\kappa_2 = 6_{-2}^{+4}$, corresponding to peak $\psi = 23\substack{+5 \\ -5}\units{^\circ}$, indicating weak but \textit{not} random spin-orbit alignment.
    }
    \label{fig:k1k2}
\end{figure*}

\begin{figure}
    \centering
    \includegraphics[width=1\linewidth]{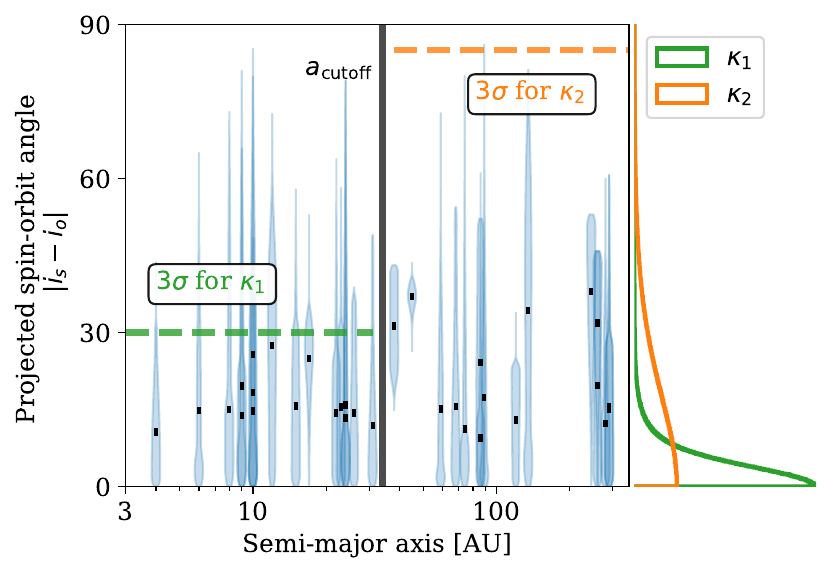}
    \caption{Projected spin-orbit angle $|i_s - i_o|$ for each star in the \JA dataset versus semi-major axis. In blue, the violin plot shows the probability distribution for each system, with black points indicating the median.
    The vertical black line separates the two populations at $a_\mathrm{cutoff}=31-38\units{AU}$. 
    Interior (exterior) to $a_\mathrm{cutoff}$, the green (orange) dotted line represents the $3\sigma$ bound of possible projected spin-orbit angles from the $\kappa_1$ ($\kappa_2$) posterior. 
    \textbf{Right:} Inferred population-level probability distributions of $|i_s - i_o|$ computed from the posterior of $\kappa_1$ and $\kappa_2$ in Fig. \ref{fig:k1k2}.
    }
    \label{fig:data_comparison}
\end{figure}

As the piecewise model with $a_\mathrm{cutoff} \sim 31-38$ AU is favored, we now study how the parameters in this model are described by sampling the posterior distributions for $\kappa_1$ and $\kappa_2$. Setting $a_\mathrm{cutoff} = 31-38 \units{AU}$, as it represents the maximal Bayes factor as shown in Fig. \ref{fig:bayes_factor}, we present the posterior distributions for $\kappa_1$ and $\kappa_2$ in Fig. \ref{fig:k1k2}.

For $\kappa_1$, the mode and 68\% HDPI is $48\substack{+194 \\ -32}$, corresponding to spin-orbit angle distributions peaking at $\psi =  8\substack{+6 \\ -4}\units{^\circ}$. For $\kappa_2$, the mode and 68\% HDPI is $6\substack{+4 \\ -2}$, corresponding to spin-orbit angle distributions peaking at $\psi = 23\substack{+5 \\ -5}\units{^\circ}$.
In addition, we find that the posterior distributions for $\kappa_1$ and $\kappa_2$ are not prior dominated.
We show the corner plot for this posterior distribution in Appendix \ref{appendix:corner}.

The $\kappa_1$ posterior favors large values, implying that spin-orbit angles favor alignment  within the semi-major axis cutoff.
In contrast, the $\kappa_2$ posterior peaks at a much lower value, indicating a distinct population characterized by a broader but non-random distribution of spin-orbit angles.
To determine if this claim is significantly affected by measurements near the cutoff (e.g, HD 113139 and HD 99028 with semi-major axes of $38$ and $45\units{AU}$), we reanalyze a subset of datapoints restricted to only $<30\units{AU}$ and $>60 \units{AU}$ in Appendix \ref{appendix:subset} and find that our result is robust.

To directly compare the inferred population-level posteriors to the \JA dataset, which directly constrains the orbital inclination $i_o$ and stellar spin axis inclination $i_s$ for individual systems, we compute the $3\sigma$ limits on what projected spin-orbit angles $|i_s - i_o|$\footnote{The choice of $i_s - i_o$ or $i_o - i_s$ is arbitrary so we take the absolute difference.} may arise from sampling the $\kappa_1$ and $\kappa_2$ posteriors. 
These limits are shown as horizontal dotted lines in Fig. \ref{fig:data_comparison}, alongside the $|i_s - i_o|$ data shown in blue as a violin plot. The $3\sigma$ limit (in green) for $\kappa_1$ encompasses the median (black dots) of all $|i_s - i_o|$ interior to the cutoff, but misses five systems exterior, which the broader $3\sigma$ limit (in orange) for $\kappa_2$ captures. On the right side of Fig. \ref{fig:data_comparison}, we display the possible $|i_s - i_o|$ given each $\kappa$ posterior. 

In summary, these comparisons highlight the existence of two distinct subpopulations separated by a semimajor axis cutoff at $31-38 \units{AU}$: the inner population is strongly aligned, while the outer population is weakly aligned, but not fully isotropic.

\subsection{Are there three populations?}\label{sec:results_double_cutoffs}

We now modify the piecewise model to have two semi-major axis cutoffs, allowing for three populations of spin-orbit angles each described by their own $\kappa$.
We refer to this model as the three population model.
This is done by modifying the piecewise condition in Eqn. \ref{eq:kappa} and the hyperprior Eqn. \ref{eqn:hyperprior}.
Informed by the two peaks in Fig. \ref{fig:bayes_factor}, we set the inner cutoff $a_\mathrm{cutoff,1} = 15-17 \units{AU}$ and the outer cutoff $a_\mathrm{cutoff,2} = 31-38 \units{AU}$.

In Appendix \ref{appendix:corner}, we present the corner plot for the posterior distribution of this three population model.
Now, we compare the three population model with the two population (piecewise) model previously developed, again using the Bayes factor.
We find that the Bayes factor is 1.2, indicating that it is currently not possible choose between the model with three or two populations.
Therefore, additional data is necessary to answer if there are more than two spin-orbit angle populations.

\section{Discussions}\label{sec:discussion}

\subsection{Comparison with \citet{Hale1994} and \citet{Justesen+Albrecht2020}}
Comparing our results to \citet{Hale1994}, we find a semi-major axis cutoff at $a_\mathrm{cutoff} = 31-38 \units{AU}$, consistent with the previously proposed cutoff at $\sim 30\units{AU}$. Furthermore, we likewise find that closer-in binaries are more aligned. However, \citet{Hale1994} found that wider binaries have projected spin-orbit angles consistent with isotropy, which we do not find. Our results disagree because neither the $\kappa_1$ nor $\kappa_2$ posterior have significant probability density for small values ($\ll1$), thus neither population favors random spin-orbit angles.

In \citet{Justesen+Albrecht2020}, the authors ask if their data is sufficient to reject either null hypotheses, that the spin-orbit angles originate from purely random (equivalent to $\kappa=0$ in our formulism) or perfectly aligned populations ($\kappa \to \infty$).
They found that neither null hypotheses can be rejected using the KS test, implying that their data is too noisy.

However, we find that the data is not pure noise, because if it is, then the posterior for $\kappa$ should be completely prior-dominated.
Yet, Fig. \ref{fig:single_kappa} in Sec. \ref{sec:results_single} shows that the $\kappa$ posterior differs noticeably from the prior, with a peak at $\kappa=11\substack{+10\\-6}$, indicating neither purely random nor perfectly aligned spin-orbit angles.
We conclude that the \JA data is sufficient to infer an informative $\kappa$ posterior using hierarchical Bayesian modeling.
Therefore, we are able to probe the data further to investigate if there may be multiple subpopulations present (Sec. \ref{sec:results_Bayes_factor}, \ref{sec:results_kappa}, \ref{sec:results_double_cutoffs}).

\subsection{Spin-orbit angles in context}

To place the spin-orbit angles of visual binaries in context, we compare them with spectroscopic binaries, astrometric binaries, and planetary systems.

At the smallest separations (orbital period $\sim 10\units{days}$), \citet{Marcussen+Albrecht2022} find that 40 out of 43 early-type and solar-type spectroscopic binaries are consistent with alignment, although some uncertainties are broad. This population is well represented by a Fisher distribution with $\kappa=6.1$ (corresponding to a $\psi$ distribution peaking at $23^\circ$), and their preference for alignment may be attributed to formation or tidal interactions. While the \JA dataset also favors alignment within $\sim 30\units{AU}$, such separations are too large for tidal interactions to play a significant role \citep{Smith+2024, Marcussen+2024}.

At intermediate separations ($100-3000\units{days}$ or separations $0.5-5\units{AU}$), \citet{Smith+2024} find that early-type astrometric binaries have a $75\%\pm5\%$ spin-orbit alignment fraction. They conclude that aligned $75\%$ population likely formed via disk fragmentation, where spin-orbit angles are expected to start nearly aligned, yet post-formation disk accretion may excite spin-orbit angles moderately. The remaining $25\%$ inconsistent with alignment instead likely formed via core fragmentation and subsequent orbital decay, which naturally misaligned spin-orbit angles. These formation pathways are similar to those expected for visual binaries in the \JA dataset, although weaker orbital decay may allow them to remain at wider separations.

At wide separations comparable to the \JA visual binary dataset, but for planetary-mass companions, \citet{Poon+2025} find that spin-orbit angles (or planetary obliquities) are consistent with isotropy, suggesting that turbulent fragmentation may be the dominant formation pathway. Our results do not find any subpopulaton with isotropic spin-orbit angles favored, indicating that either turbulent fragmentation may not be the dominant formation pathway, post-formation mechanisms may align initially misaligned visual binaries, or that the data is not sufficient to resolve visual binaries at the widest separations $\gtrsim 100\units{AU}$.

Our analysis strongly suggests that there are at least two subpopulations of visual binaries, separated by a $\sim 30\units{AU}$ cutoff. 
One potential interpretation for this cutoff is that this is the location where different formation pathways dominate: closer-in binaries form aligned in a shared protostellar disk, while wider binaries form misaligned through turbulent fragmentation.
Additionally, our analysis hints towards a possible secondary cutoff at $\sim 10-17\units{AU}$, although this claim is currently inconclusive and requires more data to verify.

\section{Conclusions}\label{sec:conclusions}
In this work, we present a hierarchical Bayesian model, capable of inferring multiple sub-populations of stellar spin-orbit angles (Sec. \ref{sec:model}).
With this framework, we find the following results by analyzing the dataset in \citet{Justesen+Albrecht2020}:
\begin{itemize}
    \item If spin-orbit angles arise from a single one population, we model that population with a Fisher distribution, yielding a concentration parameter of $\kappa=11\substack{+10\\-6}$ (Fig. \ref{fig:single_kappa}). This indicates a moderately-aligned population (in comparison, $\kappa=0$ signifies isotropy).
    \item If spin-orbit angles originate from two populations, separated at a critical cutoff, then the cutoff is at $a_\mathrm{cutoff} = 31-38 \units{AU}$. The inner (outer) population is described by a Fisher distribution with $\kappa_1 = 48_{-32}^{+194}$ ($\kappa_2 = 6_{-2}^{+4}$), shown in Fig. \ref{fig:k1k2}. This indicates that the inner population is strongly aligned, while the outer population is weakly aligned, but not isotropically distributed.
    \item We found that the two population model is strongly preferred over the single population model with a Bayes factor of 12 (Fig. \ref{fig:bayes_factor}). This suggests that there may be two separate formation pathways (e.g., disk or turbulent fragmentation) present in the visual binary sample.
    \item Although we find some indications of another cutoff at $\sim 10-17$ AU, when comparing the two versus three population model, we find a Bayes factor of 1.2. This implies that it is inconclusive to decide if there are three or more populations with current data.
\end{itemize}

Additional data is required to increase the Bayes factor for the two population versus the single population model (a Bayes factor of 100 would be required to conclusively decide between the two models). New data (e.g., rotation period from TESS photometry for stellar binaries at wide separations, astrometry from Gaia, and rotational broadening from 
high-resolution spectroscopy) will also allow the opportunity to study if there are more than two subpopulations for the spin-orbit distribution of visual binaries.

\begin{acknowledgments}
We thank the anonymous reviewer for their constructive comments that improved this work. We thank Simon Albrecht, Alan Hale, and Songhu Wang for useful discussions. We thank Joshua Speagle for useful discussions about KS statistics and hierarchical Bayesian modelling. D.P. acknowledges support from the McCray Postdoctoral Fellowship at the University of Colorado Boulder. This research has been supported by the Natural Sciences and Engineering Research Council (NSERC) Discovery Grants RGPIN-2023-05173 and RGPIN-2020-04513.

\end{acknowledgments}

\bibliography{main}{}
\bibliographystyle{aasjournal}

\appendix
\section{Sensitivity Analyses}\label{appendix:sensitivity}

\subsection{Orbital Inclination}
\begin{figure}
    \centering
    \includegraphics[width=0.6\linewidth]{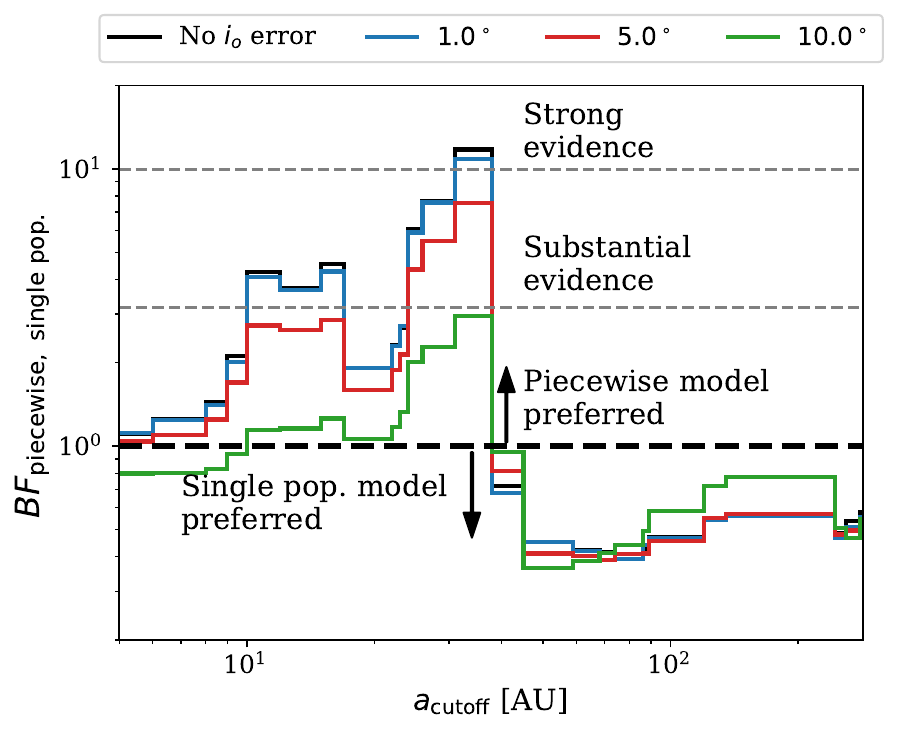}
    \caption{The Bayes factor for the piecewise (two populations) and single population model. Each curve corresponds to different assumed orbital inclination errors ($1^\circ, 5^\circ, 10^\circ$) to test the sensitivity of our Bayes factor calculations. The data in \JA assumed no orbital inclination error (black curve, same as Fig. \ref{fig:bayes_factor}). With errors included, we find that a peak remains at 31-38 AU, although they are only supported with substantial evidence.
    }
    \label{fig:orbital_inc_error}
\end{figure}

We test the sensitivity of our results to uncertainties in the orbital inclination ($i_o$), since the \JA dataset does not report uncertainties on $i_o$ (see their Table 1).
To do this, we inject random Gaussian noise into $i_o$ for each star in the sample, adopting standard deviations of $\sigma=1^\circ, 5^\circ, 10^\circ$.
We then re-perform the Bayes factor analysis in Sec. \ref{sec:results_Bayes_factor}, which is shown in Fig. \ref{fig:orbital_inc_error}.

We find that with $1^\circ$ and $5^\circ$ uncertainty, the shape of the Bayes factor curve remains consistent with the no added error curve (i.e., there remains a peak for $a_\mathrm{cutoff}$ at 31-38 AU and a secondary peak at $\sim 10-17$ AU).
The confidence in the model choice is reduced from strong to substantial evidence.
Thus, our conclusions remain the same, albeit with lower evidence.

\subsection{Stellar Binaries Near Semimajor Axis Cutoff}\label{appendix:subset}

We next test the sensitivity of our inferred $\kappa_1$ and $\kappa_2$ to potential outliers near $a_\mathrm{cutoff}=31-38\units{AU}$, where the Bayes factor drops steeply from $BF_\mathrm{piecewise, single~pop.}\sim 12$ to $<1$ at $40-50\units{AU}$ (Fig. \ref{fig:bayes_factor}).

\begin{figure}
    \centering
    \includegraphics[width=0.6\linewidth]{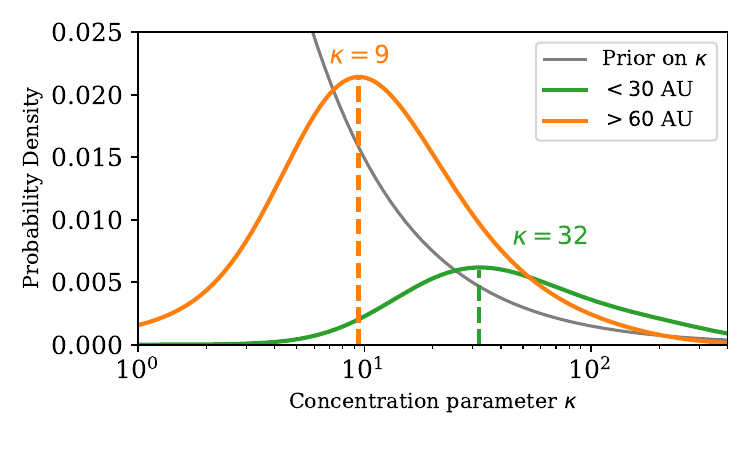}
    \caption{Posterior distributions of $\kappa$ for the $<30\units{AU}$ and $>60\units{AU}$ subsamples, showing peak values consistent with those inferred from the two-population piecewise model.}
    \label{fig:compare_two_populations}
\end{figure}

To assess whether systems near the cutoff affect our conclusion of two subpopulations, we exclude all systems with separations between $30-60 \units{AU}$.
We then apply the single-population model to each subsample (Fig. \ref{fig:compare_two_populations}). The resulting peak values of $\kappa$ remain consistent with $\kappa_1$ and $\kappa_2$ inferred from the two-population piecewise model, indicating that our conclusions are not sensitive to potential outliers near the semi-major axis cutoff.

\section{Mixture Model}\label{appendix:mixture}

We present a general model to describe a mixture of two independent stellar spin-orbit angle populations.
This is done by modifying the prior (Eqn. \ref{eq:fisher}) such that
\begin{equation}\label{eq:kappa_mixture}
    \pi(\psi|\kappa_1,\kappa_2,f) =
        f P_\mathrm{Fisher}(\kappa_1) + (1-f) P_\mathrm{Fisher}(\kappa_2),
\end{equation}
where the Fisher distribution is
\begin{equation}
    P_\mathrm{Fisher}(\kappa) = \frac{\kappa}{2\sinh{\kappa}}\exp{(\kappa \cos{\psi})}\sin{\psi}.    
\end{equation}
Like in the piecewise model, $\kappa_1$ and $\kappa_2$ describe the two populations' spin-orbit angle distribution.
The variable $f$ describes the two populations' weighting.
Note that there is a degeneracy here, since the labels $\kappa_1$ and $\kappa_2$ can be swapped to describe the same population.
We impose that $\kappa_1$ and $\kappa_2$ must be two different populations with an additional condition $\kappa_1 > \kappa_2$.
As seen, this model is independent of the stars' semi-major axis, in contrast to the other models presented in this paper.

\begin{figure}
    \centering
    \includegraphics[width=0.6\linewidth]{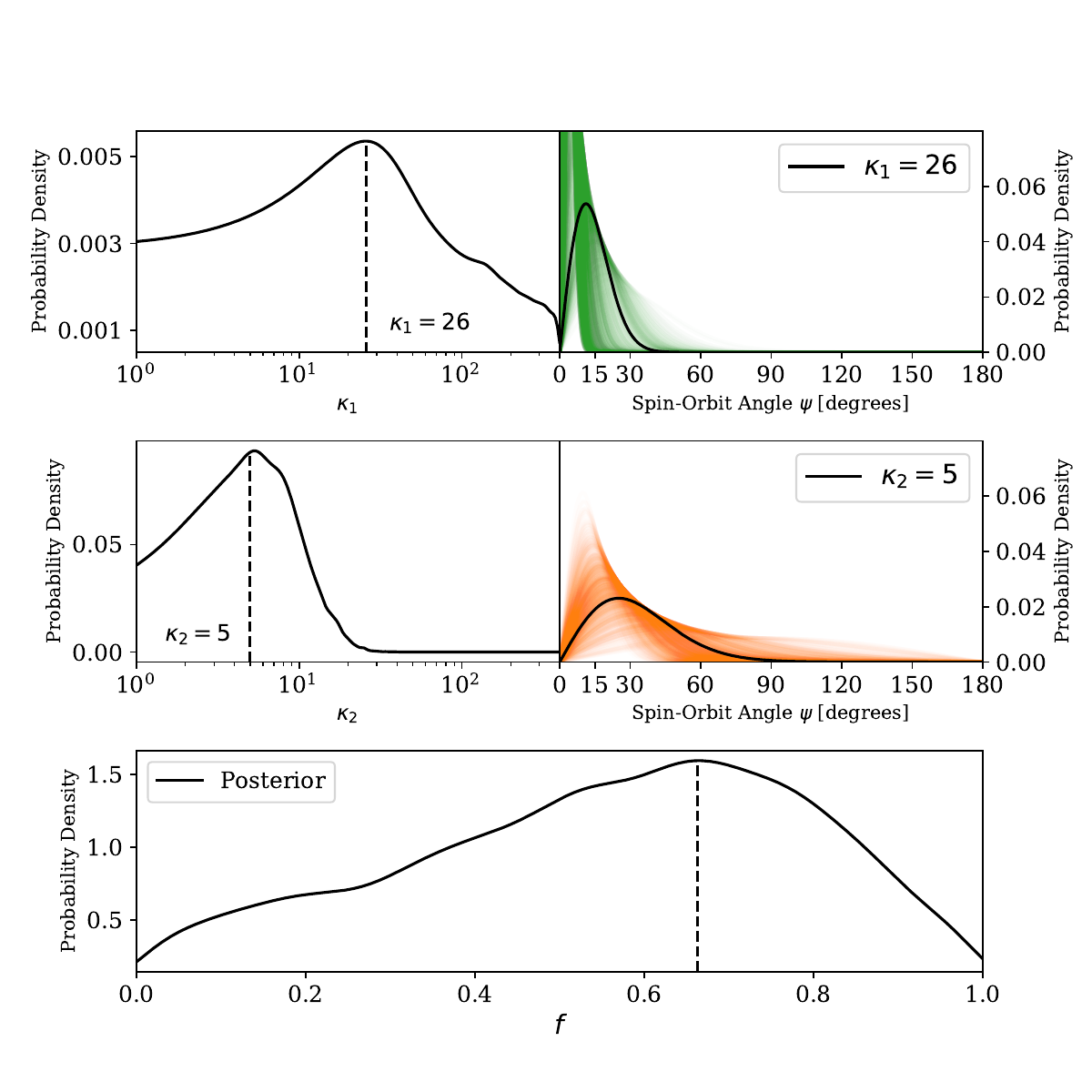}
    \caption{Posterior distributions of $\kappa_1$, $\kappa_2$, and $f$ for the mixture model. For the $\kappa$ posteriors, we also show the 1000 corresponding $\psi$ distributions on the right panel. The mode and 68\% highest density probability interval (HDPI) are $\kappa_1 = 26_{-19}^{+179}$, $\kappa_2=5_{-3}^{+5}$, and $f = 0.66_{-0.29}^{+0.21}$. This result indicates that there are two distinct stellar spin-orbit angle distributions.
}
    \label{fig:mixture_kf}
\end{figure}

In Fig. \ref{fig:mixture_kf}, we present the posterior distributions for $\kappa_1$, $\kappa_2$, and $f$ from this mixture model.
As seen, there remain two populations, as found in the piecewise model.

\section{Evidence}\label{appendix:evidence}
Here, we show the evidence used in Sec. \ref{sec:results_Bayes_factor} to calculate the Bayes factor (Figs. \ref{fig:bayes_factor}, \ref{fig:orbital_inc_error}).
The evidence for the single population model is
\begin{equation}
    Z_\mathrm{single} = \int P(\kappa) \times
  \left(\prod_{i=1}^n \int \pi(\{\psi_i, i_{o,i}\}|\kappa) ~\mathcal{L}(\{\hat{i}_{s,i},\hat{i}_{o,i}\}|\{\psi_i, i_{o,i}\}) ~\dd \psi_i ~\dd i_{o,i}\right) \dd\kappa
\label{eq:evidence_single}
\end{equation}
where $P(\kappa)$ is defined in Eqn. \ref{eq:prior}.
The evidence for the piecewise model at a given cutoff is
\begin{equation}
    Z_\mathrm{piecewise}(a_\mathrm{cutoff}) = P(a_\mathrm{cutoff}) \int P(\kappa_1)P(\kappa_2) \times \left(
  \prod_{i=1}^n \int \pi(\{\psi_i, i_{o,i}\}|\kappa) ~\mathcal{L}(\{\hat{i}_{s,i},\hat{i}_{o,i}\}|\{\psi_i, i_{o,i}\}) ~\dd \psi_i ~\dd i_{o,i}\right) \dd\kappa_1 ~\dd\kappa_2
\label{eq:evidence_piecewise}
\end{equation}
where the hyperpriors are
\begin{equation}
    P(a_\mathrm{cutoff}) = \delta(a_\mathrm{cutoff}) = \begin{cases} 
        1, & a = a_\mathrm{cutoff} \\
        0, & \mathrm{otherwise}
    \end{cases}
\end{equation}
and $P(\kappa_1)$ and $P(\kappa_2)$ take the functional form of Eqn. \ref{eq:prior}.
In the prior $\pi(\{\psi_i, i_{o,i}\}|\kappa)$, the parameter $\kappa$ is described in Eqn. \ref{eq:kappa}.

The evidence in Sec. \ref{sec:results_double_cutoffs} are derived from the same procedure.

\section{Corner plots}\label{appendix:corner}
\begin{figure}
    \centering
    \includegraphics[width=0.5\linewidth]{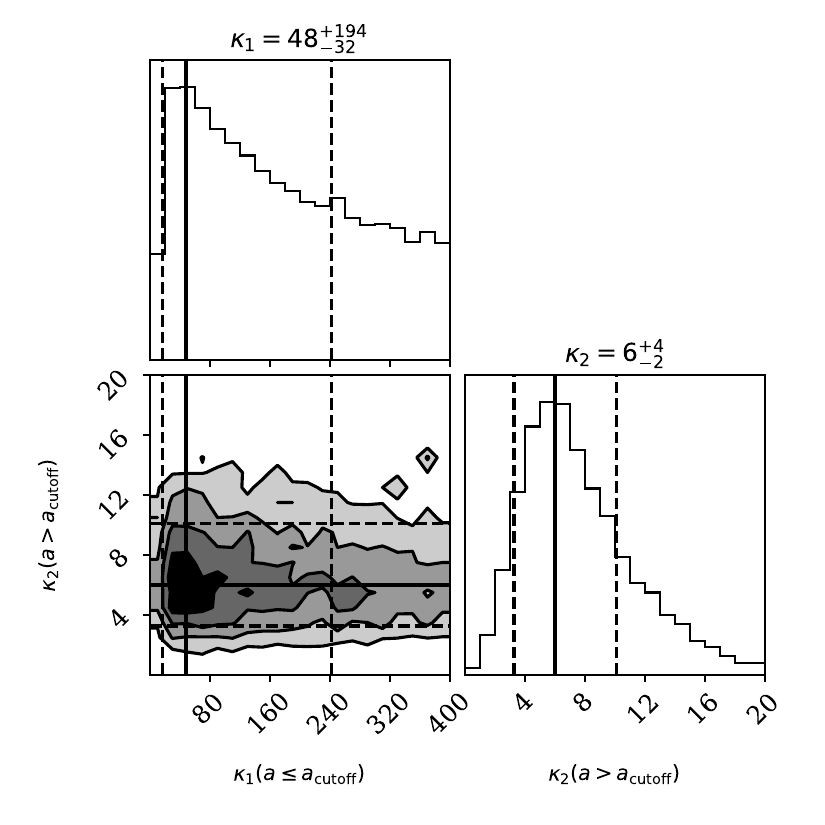}
    \caption{Corner plot for the two populations of spin-orbit angle (piecewise) model. The values shown in the figure are the mode and 68\% highest density probability interval.}
    \label{fig:corner_2pops}
\end{figure}

\begin{figure}
    \centering
    \includegraphics[width=0.7\linewidth]{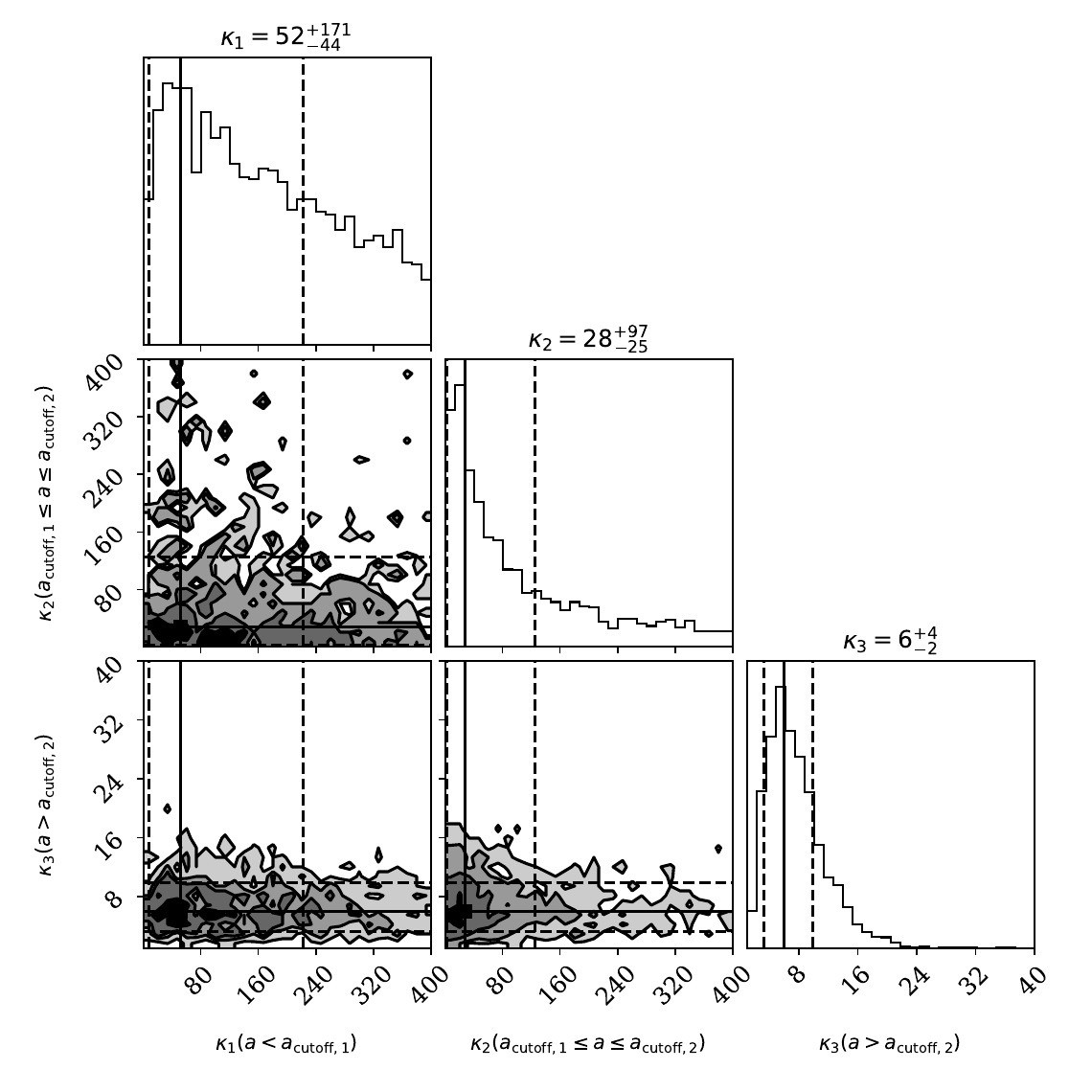}
    \caption{Corner plot for the three populations of spin-orbit angle model. The values shown in the figure are the mode and 68\% highest density probability interval.}
    \label{fig:corner_3pops}
\end{figure}

In Fig. \ref{fig:corner_2pops}, we show the corner plot for the two population (piecewise) model (see Sec. \ref{sec:results_kappa} and Fig. \ref{fig:k1k2}).
In Fig. \ref{fig:corner_3pops}, we show the corner plot for the three population model (Sec. \ref{sec:results_double_cutoffs}).
Both posteriors are sampled using \texttt{emcee}.

\end{document}